\documentclass[12 pt,twoside,aps,prd,amsmath,amssymb,
tightenlines,showpacs,showkeys,eqsecnum]{revtex4}
\usepackage{mathptmx}
\usepackage{bm}
\usepackage{graphicx}
\usepackage{hyperref}
\def \myfigures #1#2#3#4#5#6#7#8
{\begin{figure}[ht]
    \begin{center}
        \includegraphics[width=#2 \textwidth]{#1.eps}
        \hfill
        \includegraphics[width=#6 \textwidth]{#5.eps}
        \parbox[t]{#4\textwidth}{\caption {#3}\label{#1}}
        \hfill
        \parbox[t]{#8\textwidth}{\caption {#7}\label{#5}}
    \end{center}
\end{figure} }

\def\myfigure #1#2#3#4
{\begin{figure}[ht]\begin{center}
\includegraphics[width=#2 \textwidth]{#1.eps}
\parbox[t]{#4\textwidth}{\caption{#3}\label{#1}}
\end{center}\end{figure}}

\date{\today}
\begin{document}
\baselineskip -24pt
\title{Some remarks on Bianchi type-II, VIII and IX models}
\author{Bijan Saha}
\affiliation{Laboratory of Information Technologies\\
Joint Institute for Nuclear Research, Dubna\\
141980 Dubna, Moscow region, Russia} \email{bijan@jinr.ru}
\homepage{http://bijansaha.narod.ru}

\begin{abstract}

Within the scope of anisotropic non-diagonal Bianchi type-II, VIII
and IX spacetime it is shown that the off-diagonal components of the
corresponding metric impose severe restrictions on the components of
the energy momentum tensor in general. If the energy momentum tensor
is considered to be diagonal one, the spacetime, expect a partial
case of BII, becomes locally rotationally symmetric.

\end{abstract}

\keywords{Bianchi type cosmological models, Energy momentum tensor}

\pacs{98.80.Cq}

\maketitle

\bigskip

\section{Introduction}

Spatially homogeneous and anisotropic cosmological models play a
significant role in the description of the large scale behavior of
the Universe. In search of a realistic picture of the early Universe
such models have been widely studied within a framework of General
Relativity. In this note we confine our study within the scope of
Bianchi type-II (BII), type-VIII (BVIII) and type-IX (BIX)
space-times, which has recently been studied by a number of authors.
Christodoulakis {\it et al} \cite{Chris} investigated the set of
spacetime general coordinate transformations which leave the line
element of a generic Bianchi-type geometry quasi form invariant;
EinsteinТs field equations for stationary BII models with a perfect
fluid source were investigated by Nilsson and Uggla \cite{NU}; Ram
and Singh \cite{Ram} presented analytical solutions of the
Einstein-Maxwell equations for cosmological models of locally
rotationally symmetric (LRS) Bianchi type-II, VIII and IX; two-fluid
BII cosmological models were studied by Pant and Oli \cite{Pant}; a
BII cosmological model with constant deceleration parameter was
considered by Singh and Kumar \cite{Singh}; Belinchon
\cite{bel1,bel2} studied a massive cosmic string within the scope of
a BII model, while LRS BII cosmological models in the presence of a
massive cosmic string and varying cosmological constant were studied
by Pradhan {\it et. al.} \cite{Pradhan}, Kumar \cite{Kumar} and
Yadav {\it et. al.} \cite{Yadav}, respectively. Other recent work
includes exact solutions for BII cosmological model in the Jordan
Brans-Dicke scalar-tensor theory of gravitation \cite{Chauvet},
study of a BII Lyttleton-Bondi Universe \cite{Reddy1}, and
determination of an anisotropic BII cosmological model in the
presence of source-free electromagnetic fields in Lyra's manifold
\cite{Reddy2}. A general scheme for the derivation of exact solution
of Einstein equations corresponding to perfect fluid plus radiation
was outlined in \cite{Vaidya} within the scope of a BIX spacetime.
Bianchi type-VIII and IX models in the Lyttleton-Bondi universe was
studied by Shanthi and Rao \cite{Shanthi}. Bianchi type-IX cosmic
strings within the scope of scalar-tensor theory was studied in
\cite{Reddy3}. Spatially homogeneous cosmological models within the
framework of BII, BVIII and BIX are considered in \cite{Banerjee}. A
Bianchi type-IX viscous cosmological model in general relativity was
studied by Bali and Yadav \cite{Bali}. In a recent paper
\cite{SahaCEJP2011} within the scope of a BII model, we have shown
that the choice of metric and energy momentum tensors are mutually
related and off-diagonal components of Einstein tensor and energy
momentum tensor impose some restrictions on metric or material
fields. Though for the isotropic distribution of matter, the
off-diagonal components of Einstein's system of equations for BII,
BVIII and BIX models, can be ignored, it is not the case for
anisotropic distribution of matter. In this case one has to be
careful, as off-diagonal components impose some severe restrictions
on the components of energy momentum tensor or metric functions
\cite{SahaCEJP2011,Rybakov1,Rybakov2,SahaPDD69}. The main aim of
this report is to show that one has to take into account all ten
Einstein equations in order to write the correct energy-momentum
tensor for the source field. In doing so together with Bianchi
type-II metric we consider the BVIII and BIX spacetime as well.

\section{Bianchi spacetime: General remarks}

We consider the anisotropic non-diagonal Bianchi spacetime in the
form
\begin{equation}
ds^2 =  dt^2 - a_1^2(t) dx_1^2 - [h^2(x_3) a_1^2(t) + f^2(x_3)
a_2^2(t)] dx_2^2 - a_3^2(t) dx_3^2 + 2 a_1^2(t)h(x_3)dx_1 dx_2,
\label{bii-ix}
\end{equation}
with $a_1,\,a_2,\,a_3$ being the functions of $t$ and $h,\,f$
functions of $x_3$ only. Defining
\begin{equation}
\delta = - \frac{1}{f}\frac{\partial^2 f}{\partial x_3^2}
\label{delta}
\end{equation}
from \eqref{bii-ix} we find BII, BVIII and BIX models, respectively
as follows:
\begin{subequations}
\label{B289}
\begin{eqnarray}
\delta & = & 0, \qquad {\rm corresponds\,\,\,to} \quad {\rm BII}
\quad {\rm
model}, \label{BII}\\
\delta & = & -1, \quad {\rm corresponds\,\,\,to} \quad {\rm BVIII}
\quad {\rm
model}, \label{BVIII}\\
\delta & = & 1, \qquad {\rm corresponds\,\,\,to} \quad {\rm BIX}
\quad
{\rm model}. \label{BIXI}\\
\end{eqnarray}
\end{subequations}

Let us write the non-zero components of Einstein tensor explicitly
corresponding to \eqref{bii-ix}
\begin{subequations}
\label{EEgen}
\begin{eqnarray}
G_1^1 &=&  \frac{\ddot a_2}{a_2} + \frac{\ddot a_3}{a_3} +
\frac{\dot a_2}{a_2}\frac{\dot a_3}{a_3} - \frac{3}{4}\frac{a_1^2
h^{\prime 2}}{f^2 a_2^2 a_3^2} - \frac{f^{\prime\prime}}{a_3^2 f} -
\frac{1}{2}\frac{a_1^2 h}{f^2 a_2^2 a_3^2} \Bigl(h^{\prime\prime} -
\frac{h^\prime f^\prime}{f}\Bigr),
\label{g11}\\
G_2^2 &=&  \frac{\ddot a_3}{a_3} + \frac{\ddot a_1}{a_1} +
\frac{\dot a_3}{a_3}\frac{\dot a_1}{a_1} + \frac{1}{4}\frac{a_1^2
h^{\prime 2}}{f^2 a_2^2 a_3^2} +  \frac{1}{2}\frac{a_1^2 h}{f^2
a_2^2 a_3^2} \Bigl(h^{\prime\prime} - \frac{h^\prime
f^\prime}{f}\Bigr),
\label{g22}\\
G_3^3 &=&  \frac{\ddot a_1}{a_1} + \frac{\ddot a_2}{a_2} +
\frac{\dot a_1}{a_1}\frac{\dot a_2}{a_2} + \frac{1}{4}\frac{a_1^2
h^{\prime 2}}{f^2 a_2^2 a_3^2}, \label{g33}\\
G_0^0 &=&  \frac{\dot a_1}{a_1}\frac{\dot a_2}{a_2} + \frac{\dot
a_2}{a_2}\frac{\dot a_3}{a_3} + \frac{\dot a_3}{a_3}\frac{\dot
a_1}{a_1} - \frac{1}{4}\frac{a_1^2 h^{\prime 2}}{f^2 a_2^2 a_3^2} -
\frac{f^{\prime \prime}}{fa_3^2}, \label{g00}\\
G_3^0 & = & \frac{f^\prime}{f}\Bigl(\frac{\dot a_3}{a_3} -
\frac{\dot a_2}{a_2}\Bigr), \label{g03}\\
G_0^3 & = & -\frac{f^\prime}{f a_3^2}\Bigl(\frac{\dot a_3}{a_3} -
\frac{\dot a_2}{a_2}\Bigr), \label{g30}\\
G_1^2 &=&  -\frac{1}{2}\frac{a_1^2}{f^2 a_2^2 a_3^2}
\Bigl(h^{\prime\prime} - \frac{h^\prime f^\prime}{f}\Bigr),
\label{g12}\\
G_2^1 &=&   h \frac{\ddot a_1}{a_1} - h  \frac{\ddot a_2}{a_2} + h
\frac{\dot a_3}{a_3}\frac{\dot a_1}{a_1} - h \frac{\dot
a_2}{a_2}\frac{\dot a_3}{a_3} + \frac{a_1^2 h h^{\prime 2}}{f^2
a_2^2 a_3^2} + \frac{h
f^{\prime\prime}}{a_3^2 f} \nonumber\\
&-& \frac{1}{2a_3^2}\Bigl(h^{\prime\prime} - \frac{h^\prime
f^\prime}{f}\Bigr) +  \frac{1}{2}\frac{a_1^2 h^2}{f^2 a_2^2 a_3^2}
\Bigl(h^{\prime\prime} - \frac{h^\prime f^\prime}{f}\Bigr).
\label{g21}
\end{eqnarray}
\end{subequations}
It can be verified that
\begin{equation}
G_2^1 = h(G_2^2 - G_1^1) + (h^2 + f^2a_2^2/a_1^2)G_1^2. \label{grel}
\end{equation}
It should be noted that in cosmology researchers usually choose the
Einstein equations in the form
\begin{equation}
G_\mu^\nu = -\varkappa T_\mu^\nu. \label{EEg}
\end{equation}
It means that relation analogous to \eqref{grel} should be held for
the components of energy momentum tensor as well, i.e.,
\begin{equation}
T_2^1 = h(T_2^2 - T_1^1) + (h^2 + f^2a_2^2/a_1^2)T_1^2. \label{trel}
\end{equation}

Considering the energy momentum tensor in diagonal form is a common
feature in cosmology. In connection with that we assume
\begin{equation}
T_\mu^\nu = \{T_0^0,\,T_1^1,\,T_2^2,\,T_3^3\}. \label{diagemt}
\end{equation}
Under this assumption, from \eqref{trel} immediately follows
\begin{equation}
T_1^1 = T_2^2, \label{t1122}
\end{equation}
i.e., {\it if the energy momentum tensor of the material field
possesses only non-zero diagonal components, the off-diagonal
components of the Einstein tensor corresponding to the metric
\eqref{bii-ix} leads to \eqref{t1122}.}

On account of \eqref{diagemt}, \eqref{t1122} and \eqref{EEgen} the
Einstein system of equations \eqref{EEg} corresponding to
\eqref{bii-ix} now can be written as
\begin{subequations}
\label{EEgen1}
\begin{eqnarray}
\frac{\ddot a_2}{a_2} + \frac{\ddot a_3}{a_3} + \frac{\dot
a_2}{a_2}\frac{\dot a_3}{a_3} - \frac{3}{4}\frac{a_1^2 h^{\prime
2}}{f^2 a_2^2 a_3^2} + \frac{ \delta }{a_3^2}  &=&
-\varkappa T_1^1, \label{g11n}\\
\frac{\ddot a_3}{a_3} + \frac{\ddot a_1}{a_1} + \frac{\dot
a_3}{a_3}\frac{\dot a_1}{a_1} + \frac{1}{4}\frac{a_1^2 h^{\prime
2}}{f^2 a_2^2 a_3^2}  &=&
-\varkappa T_1^1, \label{g22n}\\
\frac{\ddot a_1}{a_1} + \frac{\ddot a_2}{a_2} + \frac{\dot
a_1}{a_1}\frac{\dot a_2}{a_2} + \frac{1}{4}\frac{a_1^2 h^{\prime
2}}{f^2 a_2^2 a_3^2} &=& -\varkappa T_3^3, \label{g33n}\\
\frac{\dot a_1}{a_1}\frac{\dot a_2}{a_2} + \frac{\dot
a_2}{a_2}\frac{\dot a_3}{a_3} + \frac{\dot a_3}{a_3}\frac{\dot
a_1}{a_1} - \frac{1}{4}\frac{a_1^2 h^{\prime
2}}{f^2 a_2^2 a_3^2} + \frac{\delta}{a_3^2} &=& -\varkappa T_0^0, \label{g00n}\\
\frac{f^\prime}{f}\Bigl(\frac{\dot a_2}{a_2} - \frac{\dot
a_3}{a_3}\Bigr) &=& 0, \label{g033}\\
h^{\prime\prime} - \frac{h^\prime f^\prime}{f} &=& 0,
\label{g12n}\\
\frac{\ddot a_1}{a_1} - \frac{\ddot a_2}{a_2} + \frac{\dot
a_3}{a_3}\frac{\dot a_1}{a_1} - \frac{\dot a_2}{a_2}\frac{\dot
a_3}{a_3} + \frac{a_1^2 h^{\prime 2}}{f^2 a_2^2 a_3^2}  -
\frac{\delta}{a_3^2} &=& 0. \label{g21n}
\end{eqnarray}
\end{subequations}
Here we take into account \eqref{g12n} to rewrite \eqref{g11n},
\eqref{g22n} and \eqref{g21n}. It should be noted that under the
condition \eqref{t1122} the Eq. \eqref{g21n} is identically
fulfilled, hence can be omitted.

From \eqref{g12}  we obtain
\begin{equation}
\frac{h^{\prime \prime}}{h^\prime} = \frac{f^{\prime}}{f},
\label{g12n1}
\end{equation}
with the solution
\begin{equation}
h^{\prime} = f, \label{h}
\end{equation}
where the constant is taken to be unity.

So finally we can rewrite the system \eqref{EEgen1} as follows:
\begin{subequations}
\label{EEgen2}
\begin{eqnarray}
\frac{\ddot a_2}{a_2} + \frac{\ddot a_3}{a_3} + \frac{\dot
a_2}{a_2}\frac{\dot a_3}{a_3} - \frac{3}{4}\frac{a_1^2}{a_2^2 a_3^2}
+ \frac{ \delta }{a_3^2} &=&
-\varkappa T_1^1, \label{g11n2}\\
\frac{\ddot a_3}{a_3} + \frac{\ddot a_1}{a_1} + \frac{\dot
a_3}{a_3}\frac{\dot a_1}{a_1} + \frac{1}{4}\frac{a_1^2}{a_2^2 a_3^2}
&=& -\varkappa T_1^1, \label{g22n2}\\
\frac{\ddot a_1}{a_1} + \frac{\ddot a_2}{a_2} + \frac{\dot
a_1}{a_1}\frac{\dot a_2}{a_2} + \frac{1}{4}\frac{a_1^2}{a_2^2
a_3^2} &=& - \varkappa T_3^3, \label{g33n2}\\
\frac{\dot a_1}{a_1}\frac{\dot a_2}{a_2} + \frac{\dot
a_2}{a_2}\frac{\dot a_3}{a_3} + \frac{\dot a_3}{a_3}\frac{\dot
a_1}{a_1} - \frac{1}{4}\frac{a_1^2}{a_2^2 a_3^2}
+ \frac{\delta}{a_3^2} &=& - \varkappa T_0^0, \label{g00n2}\\
\frac{f^\prime}{f}\Bigl(\frac{\dot a_2}{a_2} - \frac{\dot
a_3}{a_3}\Bigr) &=& 0. \label{g033n}
\end{eqnarray}
\end{subequations}

In what follows, we consider some concrete form of Bianchi metrics.
In doing so we first solve the equation \eqref{delta}, i.e.,
\begin{equation}
\frac{\partial^2 f}{\partial x_3^2} + \delta f = 0, \label{deltan}
\end{equation}
for different values of $\delta$.

\section{Bianchi Spacetime: Concrete examples}

In this section we study the properties system given in the
foregoing section for the concrete choice of $\delta$.

\subsection{BII metric}

The metric \eqref{bii-ix} gives rise to BII space time if $\delta =
0$,
\begin{equation}
\frac{\partial^2 f}{\partial x_3^2} = 0. \label{delbii}
\end{equation}
Eqs. \eqref{delbii} allows two cases:

{\bf Case I} As a partial solution we may consider the case when
$f^\prime = 0$. For simplicity we may write $f = 1$. Then from
\eqref{h} we obtain $h = x_3$ setting the integration constant to be
trivial. In this case from \eqref{g033} one finds that the model
allows a general case with $a_2 \ne a_3$.

The corresponding metric takes the form \cite{Pradhan,SahaCEJP2011}
\begin{equation}
ds^2 =  dt^2 - a_1^2 dx_1^2 - [x_3^2 a_1^2 +  a_2^2] dx_2^2 - a_3^2
dx_3^2 + 2 a_1^2 x_3 dx_1 dx_2. \label{bii1}
\end{equation}
The Einstein system in this case reads
\begin{subequations}
\label{EEgen2BII1}
\begin{eqnarray}
\frac{\ddot a_2}{a_2} + \frac{\ddot a_3}{a_3} + \frac{\dot
a_2}{a_2}\frac{\dot a_3}{a_3} - \frac{3}{4}\frac{a_1^2}{a_2^2 a_3^2}
&=& -\varkappa T_1^1, \label{g11n2BII1}\\
\frac{\ddot a_3}{a_3} + \frac{\ddot a_1}{a_1} + \frac{\dot
a_3}{a_3}\frac{\dot a_1}{a_1} + \frac{1}{4}\frac{a_1^2}{a_2^2 a_3^2}
&=& -\varkappa T_1^1, \label{g22n2BII1}\\
\frac{\ddot a_1}{a_1} + \frac{\ddot a_2}{a_2} + \frac{\dot
a_1}{a_1}\frac{\dot a_2}{a_2} + \frac{1}{4}\frac{a_1^2}{a_2^2
a_3^2} &=& - \varkappa T_3^3, \label{g33n2BII1}\\
\frac{\dot a_1}{a_1}\frac{\dot a_2}{a_2} + \frac{\dot
a_2}{a_2}\frac{\dot a_3}{a_3} + \frac{\dot a_3}{a_3}\frac{\dot
a_1}{a_1} - \frac{1}{4}\frac{a_1^2}{a_2^2 a_3^2}
 &=& - \varkappa T_0^0. \label{g00n2BII1}
\end{eqnarray}
\end{subequations}
Thus we see that in the case considered the off-diagonal Eqn.
\eqref{g21n} can be ignored thanks to \eqref{t1122}. And it is a
must condition. In this case anisotropy in the system can be
introduced only along $x_3$ axis, i.e., magnetic field, cosmic
string etc. should be directed along $x_3$ axis.

{\bf Case II} In this case from \eqref{delbii} one finds $f^\prime =
{\rm const.} = 1$ (say), which leads to $f = x_3$ and $h = x_3^2/2$.
In this case Eq. \eqref{g033n} leads to
\begin{equation}
\frac{\dot a_2}{a_2} - \frac{\dot a_3}{a_3} = 0. \label{bcii}
\end{equation}
Taking into account that
\begin{equation}
\frac{\ddot a_2}{a_2} = \frac{d}{dt}\Bigl(\frac{\dot a_2}{a_2}\Bigr)
+ \Bigl(\frac{\dot a_2}{a_2}\Bigr)^2 = \frac{d}{dt}\Bigl(\frac{\dot
a_3}{a_3}\Bigr) + \Bigl(\frac{\dot a_3}{a_3}\Bigr)^2 = \frac{\ddot
a_3}{a_3}, \label{a23}
\end{equation}
subtraction of \eqref{g33n2} from \eqref{g22n2} leads to
\begin{equation}
T_1^1 = T_3^3. \label{t1133}
\end{equation}
It means, in case of $f^\prime \ne 0$ the model is locally
rotationally symmetric and the matter distribution is isotropic,
i.e., $T_1^1 = T_2^2 = T_3^3$.

The corresponding metric takes the form \cite{Banerjee}
\begin{equation}
ds^2 =  dt^2 - a_1^2 dx_1^2 - [\frac{1}{4} x_3^4 a_1^2 + x_3^2
a_2^2] dx_2^2 - a_2^2 dx_3^2 +  a_1^2 x_3^2 dx_1 dx_2, \label{bii2}
\end{equation}

The Einstein system in this case reads
\begin{subequations}
\label{EEgen2BII2}
\begin{eqnarray}
2\frac{\ddot a_2}{a_2} + \frac{\dot a_2^2}{a_2^2} -
\frac{3}{4}\frac{a_1^2}{a_2^4} &=&
\varkappa p, \label{g11n2BII2}\\
\frac{\ddot a_2}{a_2} + \frac{\ddot a_1}{a_1} + \frac{\dot
a_1}{a_1}\frac{\dot a_2}{a_2} + \frac{1}{4}\frac{a_1^2}{a_2^4}  &=&
 \varkappa p, \label{g22n2BII2}\\
2\frac{\dot a_1}{a_1}\frac{\dot a_2}{a_2} + \frac{\dot a_2^2}{a_2^2}
- \frac{1}{4}\frac{a_1^2}{a_2^4}
 &=& - \varkappa \varepsilon, \label{g00n2BII2}
\end{eqnarray}
\end{subequations}
where we define $T_1^1 = T_2^2 = T_3^3 = - p$ and $ T_0^0 =
\varepsilon$. Thus we see that the model Universe can be described
by these three equations if and only if the pressure distribution is
isotropic.

\subsection{BVIII metric}

The metric \eqref{bii-ix} gives rise to BVIII space time if $\delta
= -1$,
\begin{equation}
\frac{\partial^2 f}{\partial x_3^2} - f = 0. \label{delbviii}
\end{equation}
One of the solutions to \eqref{delbviii} is $f = \sinh(x_3)$ with $h
= \cosh(x_3)$. As it was proved  for BII model with $f^\prime \ne
0$, the BVIII is also locally rotationally symmetric and its matter
distribution is isotropic, i.e., $T_1^1 = T_2^2 = T_3^3.$

The corresponding metric takes the form \cite{Banerjee}
\begin{equation}
ds^2 =  dt^2 - a_1^2 dx_1^2 - [\cosh^2(x_3) a_1^2 + \sinh^2(x_3)
a_2^2] dx_2^2 - a_2^2 dx_3^2 + 2 a_1^2 \cosh(x_3) dx_1 dx_2,
\label{bviii}
\end{equation}

The Einstein system in this case reads
\begin{subequations}
\label{EEgen2BVIII}
\begin{eqnarray}
2\frac{\ddot a_2}{a_2} + \frac{\dot a_2^2}{a_2^2} -
\frac{3}{4}\frac{a_1^2}{a_2^4} - \frac{1}{a_2^2}&=&
 \varkappa p, \label{g11n2BVIII}\\
\frac{\ddot a_2}{a_2} + \frac{\ddot a_1}{a_1} + \frac{\dot
a_1}{a_1}\frac{\dot a_2}{a_2} + \frac{1}{4}\frac{a_1^2}{a_2^4}  &=&
 \varkappa p, \label{g22n2BVIII}\\
2\frac{\dot a_1}{a_1}\frac{\dot a_2}{a_2} + \frac{\dot a_2^2}{a_2^2}
- \frac{1}{4}\frac{a_1^2}{a_2^4} - \frac{1}{a_2^2}
 &=& - \varkappa \varepsilon. \label{g00n2BVIII}
\end{eqnarray}
\end{subequations}
Thus we see that for the system to be consistent the pressure
distribution should be isotropic.

\subsection{BIX metric}

The metric \eqref{bii-ix} gives rise to BIX space time if $\delta =
1$,
\begin{equation}
\frac{\partial^2 f}{\partial x_3^2} + f = 0, \label{delbix}
\end{equation}
One of the solutions to \eqref{delbix} is $f = \sin(x_3)$ with $h =
\cos(x_3)$. It can be emphasized that like BVIII the BIX is also
locally rotationally symmetric and its matter distribution is
isotropic , i.e., $T_1^1 = T_2^2 = T_3^3.$

The corresponding metric takes the form \cite{Bali,Reddy3,Banerjee}
\begin{equation}
ds^2 =  dt^2 - a_1^2 dx_1^2 - [\cos^2(x_3) a_1^2 + \sin^2(x_3)
a_2^2] dx_2^2 - a_2^2 dx_3^2 + 2 a_1^2 \cos(x_3) dx_1 dx_2,
\label{bix}
\end{equation}

The Einstein system in this case reads
\begin{subequations}
\label{EEgen2BIX}
\begin{eqnarray}
2\frac{\ddot a_2}{a_2} + \frac{\dot a_2^2}{a_2^2} -
\frac{3}{4}\frac{a_1^2}{a_2^4} + \frac{1}{a_2^2}&=&
\varkappa p, \label{g11n2BIX}\\
\frac{\ddot a_2}{a_2} + \frac{\ddot a_1}{a_1} + \frac{\dot
a_1}{a_1}\frac{\dot a_2}{a_2} + \frac{1}{4}\frac{a_1^2}{a_2^4}  &=&
\varkappa p, \label{g22n2BIX}\\
2\frac{\dot a_1}{a_1}\frac{\dot a_2}{a_2} + \frac{\dot a_2^2}{a_2^2}
- \frac{1}{4}\frac{a_1^2}{a_2^4} + \frac{1}{a_2^2} &=& -\varkappa
\varepsilon. \label{g00n2BIX}
\end{eqnarray}
\end{subequations}
As in previous cases, consideration of \eqref{EEgen2BIX} imposes the
isotropic distribution of matter, otherwise the system becomes
inconsistent.

\section{conclusion}

Within the scope of anisotropic non-diagonal Bianchi type-II, VIII
and IX spacetime we study the role of the off-diagonal components of
the corresponding metric that it plays on the choice of  the energy
momentum tensor in general. It is shown that

{\it if the energy momentum tensor of the source field possesses
only non-zero diagonal components:}
\begin{equation}
T_\mu^\nu = \{T_0^0,\,T_i^i,\,T_j^j,\,T_k^k\} \label{emtijk}
\end{equation}
{\it and the space-time is given by the metric}
\begin{equation}
ds^2 =  dt^2 - a_i^2(t) dx_i^2 - [h^2(x_k) a_i^2(t) + f^2(x_k)
a_j^2(t)] dx_j^2 - a_k^2(t) dx_k^2 + 2 a_i^2(t)h(x_k)dx_i dx_j,
\label{bii-ix-ijk}
\end{equation}
{\it with $a_i,\,a_j,\,a_k$ being the functions of $t$ and $h,\,f$
functions of $x_k$ only, the off-diagonal components of the Einstein
tensor immediately lead to}
\begin{equation}
T_i^i = T_j^j. \label{tij}
\end{equation}
{\it In case of $f^\prime = 0$, which is a partial case of BII
space-time, the matter distribution may be anisotropic with magnetic
field, cosmic string etc. directed along $x_k$ axis. In case of
$f^\prime \ne 0$, the metric functions $a_j$ and $a_k$ are linearly
dependent which gives rise to locally rotationally symmetric (LRS)
Bianchi models. Moreover, in this case the off-diagonal components
of Einstein tensor leads to the isotropic distribution of matter,
i.e., the components of the energy momentum tensor must satisfy}

\begin{equation}
T_i^i = T_j^j = T_k^k. \label{tijk}
\end{equation}

{\it If the energy momentum tensor possesses non-zero off-diagonal
components, for the system to be consistent, the relation

\begin{equation}
T_j^i = h(T_j^j - T_i^i) + (h^2 + f^2a_j^2/a_i^2)T_i^j.
\label{trelij}
\end{equation}
must hold.}

In this note we give some general remarks on the choice of energy
momentum tensor. We plan to study the evolution of the Universe
given by Bianchi models filled with spinor, scalar and other fields
in near future.

\end{document}